\documentclass[amssymb,prb,twocolumn,superscriptaddress,floats]{revtex4}

\usepackage[T1]{fontenc}
\usepackage{bm}
\usepackage{graphicx}% Include figure files

\begin{document}

\title{Single-photon emission from the natural quantum dots in the
InAs/GaAs wetting layer}

\author{T. Kazimierczuk}
\author{A. Golnik}
\author{P. Kossacki}
\author{J. A. Gaj}
\thanks{deceased}
\affiliation{Institute of Experimental Physics, Faculty of Physics,
University of Warsaw, ul. Ho\.za 69, 00-681 Warszawa, Poland}
\author{Z. R. Wasilewski}
\affiliation{Institute for Microstructural Sciences, National Research
Council, Ottawa, Ontario K1A 0R6, Canada}
\author{A. Babi\'nski}
\email{Adam.Babinski@fuw.edu.pl}
\affiliation{Institute of Experimental Physics, Faculty of Physics,
University of Warsaw, ul. Ho\.za 69, 00-681 Warszawa, Poland}
\date{\today}

\begin{abstract}
Time-resolved microphotoluminescence study is presented for quantum dots
which are formed in the InAs/GaAs wetting layer. These dots are due to
fluctuations of In composition in the wetting layer. They  show spectrally
sharp luminescence lines with a low spatial density. 
We identify lines related to neutral exciton and biexciton as well as
trions. Exciton emission antibunching (second order correlation value of
$g^{(2)}(0)=0.16$) and biexciton-exciton emission cascade prove
non-classical emission from the dots and confirm their potential as single
photon sources. 
\end{abstract}


\maketitle

\section{\label{sec:Intro} Introduction}

Quantum confinement of carriers in semiconductor quantum dots (QDs) leads
to numerous effects of fundamental character. This makes them objects of
intense studies [for review see Ref. \onlinecite{optics_qd_qw}]. The
research is driven by both scientific curiosity and novel applications e.g.
in optical quantum devices based on single photon emission
\cite{santori-prb-2001}. Such photoluminescence relies on cascade process
with a spectrally distributed  multiexcitonic  emissions and recombination
of a single exciton as a final step of the sequence. Although the single
photon emission has been observed in several physical systems, its
QDs-realization promises an easy incorporation in optoelectronic
semiconductor devices. The single photon emission in QDs has been shown in
several systems including e.g. InAs/GaAs self-assembled QDs
\cite{santori-prb-2001}, CdSe/Zn(S,Se)  \cite{sebald-apl-2002},  CdTe/ZnTe
QDs \cite{couteau-apl-2004,suffczynski-prb-2006} and \emph{naturally}
occurring QDs with confinement resulting from monolayer width fluctuations
of a thin GaAs/GaAlAs \cite{hours-apl-2003} quantum well. In this work we
investigate a potential of \emph{natural} QDs present in a wetting layer
(WLQDs), which accompanies InAs/GaAs self-assembled QDs
\cite{babinski-apl-2008} as single photon emitters. We report on
correlation spectroscopy of excitonic complexes confined in the dots.
Experimental evidence of the radiative cascade between the biexcitonic and
excitonic emission in the WLQDs is presented. Charged exciton-neutral
exciton cross-correlation experiments confirm attribution of the excitons
to the same WLQD. Radiative lifetimes of the excitons have been measured
and found to be of the same order as values usually observed in InAs/GaAs
self-assembled QDs. 
The paper is organized as follows. Information on the investigated sample
and experimental setup is provided in Sec. \ref{sec:setup}. Results of
microphotoluminescence ($\mu$PL) characterization of the sample are
presented in Sec. \ref{sec:pl}. Time-resolved measurements are described in
Sec. \ref{sec:trpl}.

\section{Sample and experimental setup \label{sec:setup}}
The sample investigated in this work was grown by molecular beam epitaxy
using In-flush technique \cite{wasilewski-growth-1999}. It contains a
single layer of InAs QDs grown at 524 $^\circ$C, deposited on  GaAs
substrate covered by 800 nm GaAs buffer layer. The sample was capped with
100 nm GaAs top layer. Indium-flush was applied to the QDs at 5 nm. 

\begin{figure}
\includegraphics[width=85mm]{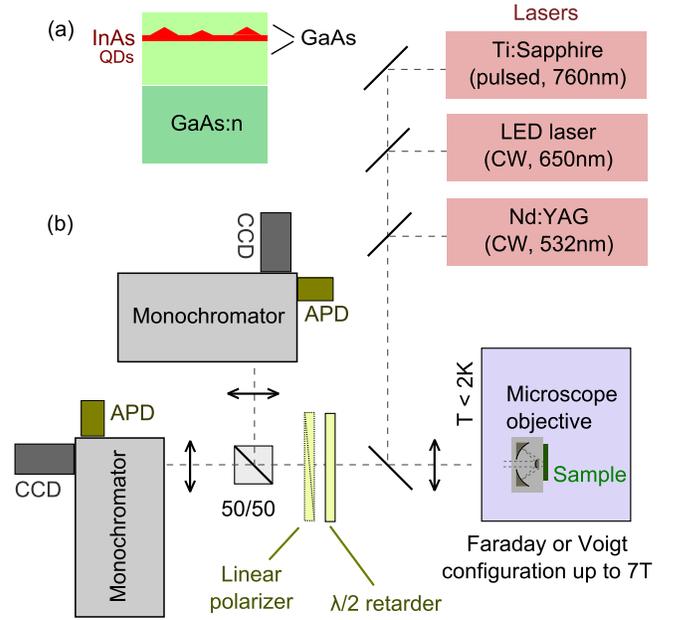}
\caption{(a) Structure of the sample. (b) Schematic of experimental setup
used in the study. \label{fig:setup}}
\end{figure}

The samples were investigated in a $\mu$PL setup presented schematically
in Fig. \ref{fig:setup}(b). All measurements were performed at low
temperature with the sample immersed in superfluid helium ($T<2$K) inside
an Oxford SpectroMag bath cryostat . The cryostat was equipped with
superconducting coils producing magnetic field up to 7T either in Faraday
or Voigt configuration. High spatial resolution was assured in the
experiment by using specially design reflection-type microscope
objective\cite{jasny} immersed together with the sample inside the
cryostat. The diameter of the laser spot on the sample surface was below
1$\mu$m.
Three different lasers were employed to excite the sample: non-resonant CW
Nd:YAG (532nm) and LED laser (650nm) or quasi-resonant femtosecond
oscillator tuned at central wavelength of 760nm.

The photoluminescence was resolved using a 0.5m diffraction spectrometer
with a CCD camera and an avalanche photodiode (APD) for ultra-fast single
photon detection. Two different time-resolved techniques were used.
Measurements of radiative lifetime were carried out using a low jitter APD
(idQuantique id50). The overall temporal resolution in this measurement was
up to 40 ps.

The other time-resolved technique used in our experiment was a
single-photon correlation measurement.
In this case two independent monochromators with high efficiency
Perkin-Elmer APDs were used. The monochromators were tuned to pass photons
from a single excitonic transition, set independently for each
spectrometer. The setup was arranged in
Hanbury-Brown--Twiss\cite{hbt_setup} configuration with a 50/50
non-polarizing beamsplitter. Signals from the APDs were recorded using
multichannel picosecond event timer HydraHarp400. The timer was started by
a photon detection in one APD and stopped by a photon detection in the
other APD. A histogram of the events was built as a function
of time delay between detection of two photons. Total temporal resolution
of the setup was estimated as 1.1 ns. Accumulation time of a single
correlation histogram was up to several hours, depending on the intensity
of investigated emission lines.

\section{Photoluminescence results \label{sec:pl}}

\begin{figure}
\includegraphics{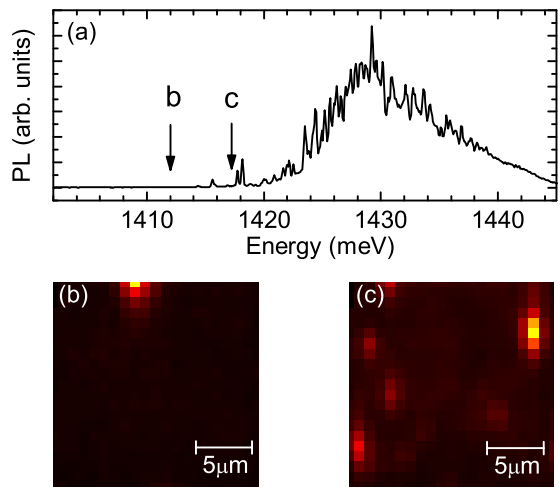}
\caption{(a) The $\mu$PL spectrum from the structure with WLQDs. (b-c)
Spatial mapping of
PL intensity monitored at (b) 1412 meV and (c) 1417 meV. The range of the
presented maps was equal to 20$\times$20 $\mu$m$^2$ \label{fig:pl}}
\end{figure}

The PL spectrum from the investigated structure consists of the emission
related to the GaAs barrier (1.49 eV), self-assembled
QDs\cite{babinski-prb-2006} (1.25--1.30 eV), and the emission related to
the WL (1.42--1.44 eV). The $\mu$PL spectrum in the energy range of the
expected WL-emission consists of a broad peak (Fig. \ref{fig:pl}(a)) with
well-resolved sharp line structure. This PL has been attributed to
recombination of electrons and heavy holes in the WL observed in similar
QDs systems \cite{sanguinetti-prb-1999}. The structure of the peak shows
that the WL is strongly disordered due to composition and strain
fluctuations. The disorder leads to potential fluctuations, which localize
excitons. At some spots on the sample several well-resolved lines emerge in
the spectrum at lower (by up to 10 meV) energies. We attribute these
emission lines from the low-energy tail of the WL-related band as being
related to recombination of excitons in the WLQDs. 

In order to provide more comprehensive characterization of the WLQD
emission we analyzed PL properties of several different dots. All of them
were characterized by relatively low emission energy, which was necessary
to resolve the QD signal from the much stronger WL emission. We found that
different dots shared the same PL pattern, including the same spectral
order and similar spacing between various PL lines. 

\begin{figure}
\includegraphics[width=85mm]{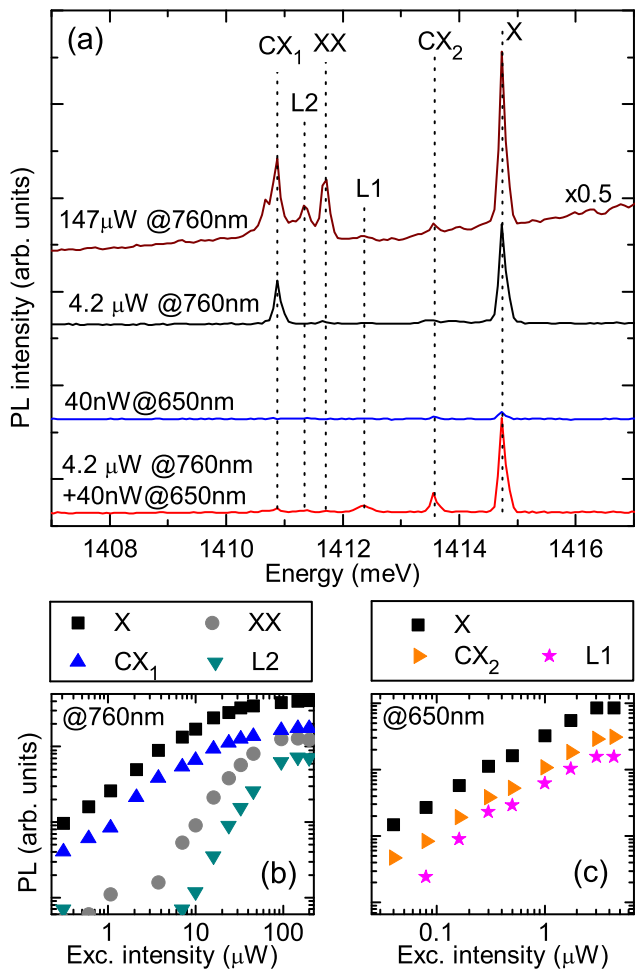}
\caption{(a) PL spectrum of the same single QD under different excitation
conditions. 
(b-c) Dependence of the PL intensity on excitation power under excitation
at (b) pulsed 760nm and (c) CW 650nm. \label{fig:micropl}}
\end{figure}

Typical spectrum of a single WLQD is presented in Fig.
\ref{fig:micropl}(a). Intensities of various PL lines present in the
spectrum depend on parameters of excitation. With a weak pulsed excitation
(i.e. far from saturation of the QD emission) at 760 nm (1.63 eV) the PL
spectrum features two main emission lines related to a neutral exciton (X)
and charged exciton (CX$_1$) recombination. Under stronger excitation
additional lines emerge: a line related to a neutral biexciton (XX) and a
few other weaker lines denoted for the sake of present study as L1 and L2.
Detailed identification of these transitions is beyond the scope of the
present study.

\begin{figure}
\includegraphics[width=85mm]{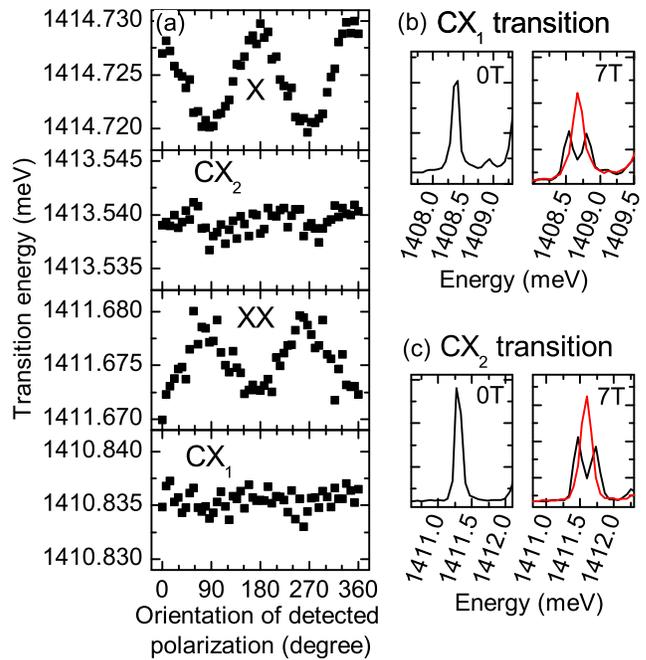}
\caption{(color online) 
(a) Transition energies measured in selected linear polarization.
Oscillatory behavior
indicates small anisotropic splitting of about 9 $\mu$eV.
Plots in the right-hand side panels present splitting pattern of (b)
CX$_1$ and (c) CX$_2$ transition
in magnetic field (Voigt configuration) measured for another dot. Black
and red (grey) curves correspond to two perpendicular linear polarizations.
\label{fig:ident}}
\end{figure}

The identification of transitions X, XX, CX$_1$, CX$_2$ is firmly
supported by measurements of polarization-resolved PL and
magnetophotoluminescence. In the former case we recorded PL spectra while
changing the orientation of detected linear polarization. By fitting 
gaussian profiles to the analyzed spectral lines we extracted apparent
energies of each transition. The energies of X and XX transitions exhibited
clear sine-like oscillations (Fig. \ref{fig:ident}(a)) evidencing a
splitting into two linearly polarized components\cite{bester-prb-2003} not
resolved within our experimental resolution. The splitting is related to
anisotropic exchange electron-hole interaction, which is characteristic of
neutral excitons in QDs\cite{gammon-prl-1996}. The splitting pattern allows
to attribute the X and XX emission lines to the neutral exciton and
biexciton respectively.
The value of anisotropic splitting for the dot presented in Fig.
\ref{fig:ident}(a) yielded $9\pm2$ $\mu$eV.  The relatively low anisotropy
of the neutral
exciton is similar to the values found in natural QDs formed by interface
fluctuations in thin quantum wells \cite{gammon-prl-1996}.

Another important parameter of the excitation was the wavelength of the
laser line. As we show in Fig. \ref{fig:micropl} WLQD emission can be
strongly affected by this parameter, plausibly due to charge redistribution
induced by higher-energy excitation, e.g. at 650nm (1.91 eV). Data in the
Figure were obtained with
relatively weak excitation of higher energy --- 100 times lower than
simultaneous low-energy excitation --- evidencing the sensitivity of the PL
spectrum to the induced charge redistribution. The main observed change is
quenching of a charged exciton line
CX$_1$ accompanied with emerging of another line recognized as a charged
exciton of the opposite sign CX$_2$.

No effect of in-plane anisotropy was detected for lines CX$_1$ and CX$_2$,
which confirms their attribution to charged excitons. 
Observation of both the neutral and the charged excitons is usually
reported in spectroscopic studies of nonresonantly excited QDs 
\cite{finley-prb-2001}  Further confirmation of the charged-exciton lines
identification was provided by measurements of PL in the magnetic field in
Voigt configuration (Fig. \ref{fig:ident}(b-c)). In such a configuration, a
splitting into four components is expected for charged exciton transition
as opposed to two bright components for the neutral exciton
\cite{bayer-prb-2002}. Our spectroscopic data confirm such a prediction for
both CX$_1$ and CX$_2$ transitions (Fig.\ref{fig:ident}(b) and
Fig.\ref{fig:ident}(c) respectively).
In both cases we observe symmetrical splitting into two pairs of lines
linearly polarized along or perpendicularly to the field direction. The
splitting of the inner doublet at 7T is observable only as an increase in
the linewidth while the splitting of the outer doublet is more pronounced
(0.29 meV at 7T as shown in Fig. \ref{fig:ident}(b) and 0.27 meV in case of
Fig. \ref{fig:ident}(c)).

\section{Time resolved measurements \label{sec:trpl}}

\begin{figure}
\includegraphics{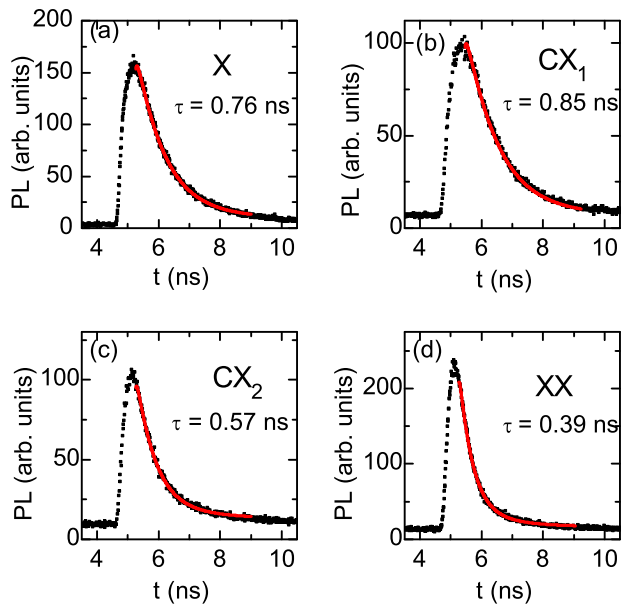}
\caption{Time-dependence of PL signal for various lines of a single WLQD:
(a) X line, (b) CX$_1$ line, (c) CX$_2$ line, and (d) XX line. 
Solid lines demonstrate fits of single exponential decays convolved
with the response function of the detection setup.
\label{fig:lifetimes}}
\end{figure}

In order to verify feasibility of using WLQDs as non-classical light 
emitters we investigated dynamical properties of WLQD PL by measurements
of 
time-resolved photoluminescence  and single photon correlations.

The results of the measurements revealed sub-nanosecond life-times
of the excitons confined in the WLQDs (Fig. \ref{fig:lifetimes}). For
single
excitons we found life-times of 0.57 to 0.85 ns depending on the charge
state.
The neutral biexciton XX line exhibited shorter life-time of 0.39 ns.
Measured life-times are two times shorter than life-time measured for an
array of self-assembled InAs/GaAs QDs in the same sample which yields about
1.6 ns\cite{auer-unpublished}.
Short life-time of the WLQDs may be related to the weak lateral
confinement\cite{hours-aip-2006}. 
Effective carrier recombination in the WLQDs may have important
implications for the dynamics of carrier trapping in self-assembled QDs in
optoelectronic structures.  Trapping electrons and holes in the WLQDs may
influence lateral carrier transport in the WL \cite{heitz-prl-1997,
lobo-prb-1999, moskalenko-prb-2006}. A decreased efficiency of QDs feeding
with carriers may affect performance of optoelectronic devices. Moreover, a
possible in-plane energy transfer to lower energy
dots\cite{kazimierczuk-prb-2009} would also have detrimental effect on
devices based on self-assembled dots due to uncontrolled re-excitation from
WLQD system.

\begin{figure}
\includegraphics{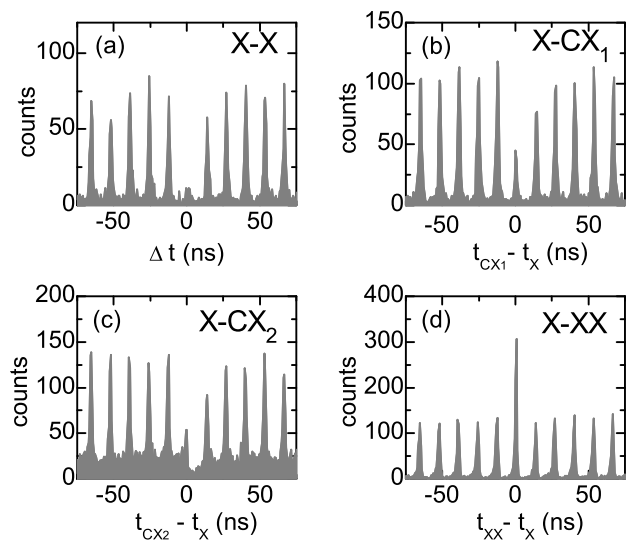}
\caption{Correlations between photons emitted from a single WLQD
(a) autocorrelation X-X, (b) crosscorrelation X-CX$_1$ line, 
(c) crosscorrelation X-CX$_2$, and 
(d) crosscorrelation X-XX. Cross-correlation between X and CX$_2$ lines
was measured with additional weak CW illumination.
\label{fig:correlations}}
\end{figure}

The ultimate demonstration of the QDs single-photon emission was provided
by means of single-photon correlation measurements. Fig.
\ref{fig:correlations}
presents a set of correlation results as histograms of number of detected
photon pairs of given temporal separation. All measured histograms consist
of 
distinct peaks separated by laser repetition period of 13.2 ns. The width
of 
each peak is governed by the relaxation time addressed in time resolved PL
measurements 
and the temporal resolution of the setup.

Autocorrelation of the X line shown in Fig. \ref{fig:correlations}
exhibits well
pronounced suppression of a central ($\Delta t=0$) peak related to a
photon 
antibunching effect. The relative height
of the central peak amounted to second order correlation parameter of 
$g^{(2)}(t\!=\!0) = 0.16\pm 0.07$. Comparison of obtained
$g^{(2)}(t\!=\!0)$ parameter with its 
threshold value of 0.5 \cite{michler-science-2000} unequivocally proves
the single-photon character of light emitted due to excitonic recombination
in the studied WLQD.

We exploited two independent monochromators in the Hanbury-Brown--Twiss
configuration to measure 
also cross-correlations between different lines of a single WLQD 
(Fig. \ref{fig:correlations}(b-d)). Strong correlation evidenced as
increase
or decrease of the intensity of the central peak confirmed our previous 
attribution of PL lines to the same single WLQD. 

Each cross-correlation histogram reflects the relation between
correlated transitions. Specifically, in the case of cross-correlations
between
neutral and charged excitons we observed an antibunching effect comparable
to the
X-X autocorrelation discussed previously. A faint asymmetry of a side peak
height
may be related to processes of a single carrier capture
\cite{suffczynski-prb-2006},
however low signal intensity did not allow us to elaborate on this effect.

An interesting effect was observed for X-CX$_2$ cross-correlation (Fig.
\ref{fig:correlations}(c)), which was measured using weak above-barrier
illumination to increase the intensity of CX$_2$ line. The mixed excitation
regime (pulsed excitation + CW illumination) was reflected in correlation
histogram, which featured both well-pronounced peaks at laser repetition
period and recognizable CW background. The height of the peaks exhibited
similar asymmetry to the one observed in case of X-CX$_1$ correlation and
attributed to the single carrier capture
mechanism\cite{suffczynski-prb-2006}. A similar effect was observed also
for the CW component of the correlation histogram. The photon count in the
range between the central peak and the consecutive one was found much
smaller then in the other sections of the histogram. The difference is
easily interpreted in terms of the single carrier mechanism invoked
earlier. Directly after the recombination of the neutral exciton the dot is
empty and the emission of the charged exciton require capturing of the
three carriers. Such a process is unlikely to occur using only the weak
illumination. Conversely, the next laser pulse can excite the dot to the
dark exciton state and thus increase the probability of the creation of
charged exciton by the illumination during the next inter-pulse period. The
difference between these two scenarios (3 carrier vs 1 carrier capture)
accounts for the discussed effect in the correlation histogram.

Finally, we also demonstrated a cascade-type cross-correlation between XX
and X lines,
as shown in Fig. \ref{fig:correlations}(d). In this case the central peak
was more pronounced than the average peak by a factor of
$g^{(2)}(t\!=\!0)=2.5\pm 0.2$.
Obtained value well corresponds to relatively strong excitation regime
that was 
used in our experiment.

\section{Summary \label{sec:summary}}

In conclusion, we have shown that the wetting layer quantum dots
coexisting with self-assembled QDs can be used as a source of quantum
light. We were able to determine basic properties of WLQD emission,
including identification of all main PL lines. The results supporting
presented identification include PL experiments with the external magnetic
field and polarization-resolved PL measurements. Fine structure splitting
of the excitonic transition was found in range around 10$\mu$eV without any
special processing of the sample.

We also successfully demonstrated a variety of single photon correlations
between emission
lines of a single WLQD. Our results unequivocally prove that observed
emission lines originate from a single dot and further support their
identification by asymmetry of the cross-correlation histograms. Moreover,
our demonstration of biexciton-exciton cascade in this system opens a
possibility to pursue entangled photon pair generation from WLQDs by
further reduction of their in-plane anisotropy.

\begin{acknowledgments}
The work has been supported in part by the MTKD-CT-2005-029671, Polish
Funds for Science 2009-2011, and Foundation for Polish Science.
\end{acknowledgments}

\end{document}